\documentclass[aps,prd,twocolumn, nofootinbib,preprintnumbers]{revtex4}
\usepackage{graphicx,graphics}
\usepackage{amssymb}
\usepackage{amsmath}
\usepackage{pifont}
\usepackage{natbib}
\usepackage{subfigure,epsfig,epstopdf}
\usepackage{color}
\usepackage{bm}

\newcommand{\be}{\begin{equation}}
\newcommand{\ee}{\end{equation}}
\newcommand{\bea}{\begin{eqnarray}}
\newcommand{\eea}{\end{eqnarray}}
\newcommand{\mnras}{{Mon.Not.Roy.Astron.Soc.}}

\begin{document}
\preprint{}

\title{Electromagnetic signals from bare strange stars}

\author{Massimo Mannarelli}
\email{massimo@lngs.infn.it}
\affiliation{INFN, Laboratori Nazionali del Gran Sasso, Via G. Acitelli, 22, I-67100 Assergi (AQ), Italy}
\author{Giulia Pagliaroli}
\email{giulia.pagliaroli@lngs.infn.it}
\affiliation{INFN, Laboratori Nazionali del Gran Sasso, Via G. Acitelli, 22, I-67100 Assergi (AQ), Italy}
\author{Alessandro Parisi}
\email{alessandro.parisi@lngs.infn.it}
\affiliation{Dipartimento di Scienze Fisiche e Chimiche, Universit\`a di L'Aquila, I-67010 L'Aquila, Italy\\ INFN, Laboratori Nazionali del Gran Sasso, Via G. Acitelli, 22, I-67100 Assergi (AQ), Italy}
\author{Luigi Pilo}
\email{luigi.pilo@aquila.infn.it}
\affiliation{Dipartimento di Scienze Fisiche e Chimiche, Universit\`a di L'Aquila, I-67010 L'Aquila, Italy\\ INFN, Laboratori Nazionali del Gran Sasso, Via G. Acitelli, 22, I-67100 Assergi (AQ), Italy}

\begin{abstract}
The  crystalline color superconducting phase is believed to be the ground state of deconfined quark matter for sufficiently large values of the strange quark mass.  This phase has the remarkable property of being more rigid than any known material. It can therefore   sustain large shear stresses, supporting  torsional oscillations of large amplitude. The torsional oscillations  could lead to  observable electromagnetic  signals if  strange stars  have a crystalline color superconducting crust.  Indeed,  considering a simple model of strange star with a bare quark matter surface, it turns out that a positive charge is localized in a narrow shell about ten  Fermi thick  beneath the star surface. The electrons needed to neutralize the positive charge of quarks spill in the star exterior forming an electromagnetically bounded atmosphere hundreds of Fermi thick.  When a torsional oscillation is excited, for example by a stellar glitch, the positive charge oscillates with typical kHz frequencies, for a   crust thickness of about one-tenth of  the stellar radius, to hundreds of Hz, for a  crust thickness of  about nine-tenths of the stellar radius. Higher frequencies, of the order of few GHz, can be reached if the star crust is of the order of few centimeters thick.  We estimate the emitted power considering  emission by an oscillating magnetic dipole, finding that it can be quite large, of the order of  $10^{45}$ erg/s for a thin crust. The associated relaxation times are very uncertain, with values ranging between  microseconds and minutes, depending on the crust thickness. The radiated photons will be in part absorbed by the electronic atmosphere, but a sizable fraction of them should be emitted by the star.
\end{abstract}

\maketitle
\section{Introduction}
One of the routes for studying the properties of matter at very high densities is by the inspection of the properties of compact stellar objects (CSOs). These are stars having a mass of $1-2 M_\odot$ and a radius of about $10$
km, typically observed as pulsars. Baryonic matter inside a CSO is squeezed at densities about a factor $3$-$5$ larger than in heavy nuclei. From a  simple geometrical reasoning one can argue that in these conditions baryons are likely to lose their identity~\cite{Weber-book} and a new form of  matter should be realized.

One possibility is that the extremely high densities and low
temperatures may favor the transition from nuclear matter to
deconfined quark matter in the core of the CSO~\cite{Ivanenko1965, Ivanenko:1969gs, Collins,
  Baym:1976yu}. In this case compact (hybrid) stars featuring quark cores and a crust of standard nuclear matter would exist. 

A second possibility is that strange matter is the ground state of the
hadrons~\cite{Witten:1984rs}. In this case at high densities there should exist the possibility of converting nuclear matter to deconfined matter. The resulting CSO would be a strange
star~\cite{Alcock:1986hz, Haensel:1986qb}, {\it i.e} a CSO completely constituted of deconfined matter, see~\cite{Madsen:1998uh} for a review. 

Unfortunately these two possibilities cannot be checked by first principle calculations. Indeed at the densities relevant for CSOs, quantum chromodynamics  (QCD) is nonperturbative, because the typical  energy scale is about $\Lambda_\text{QCD}$. Moreover, lattice QCD simulations at large baryonic densities are unfeasible because of the so-called sign problem~\cite{Barbour:1986jf}, see~\cite{Aarts:2013naa} for a recent review and~\cite{Yamamoto:2014lia} for a study of an inhomogeneous phase.  

Although not firmly established by first principles, it is reasonable to expect that if deconfined quark matter is present, it should be in a color superconducting (CSC) phase~\cite{Rajagopal:2000wf, Alford:2007xm, Anglani:2013gfu}. The reason is that   the critical temperature 
of color superconductors is large, $T_{c} \simeq 0.57 \Delta$, where   
$\Delta\sim 5 -100$ MeV is the gap parameter. For the greatest part 
of the CSO lifetime, the temperature is much lower than this critical
temperature and the CSC phase is thermodynamically favored. 

It is widely accepted that at asymptotic densities, when the up, down and strange quarks  
can be treated as massless,  the color-flavor locked (CFL)
phase~\cite{Alford:1998mk} is the ground state of matter. This phase is energetically favored 
because quarks of all flavors and of all
colors form standard Cooper pairs, thus maximizing the free energy gain. 
However, considering realistic conditions realizable within CSOs a
different CSC phase could be realized. The reason is that the nonzero
and possibly large value of the strange quark mass, $M_s$, 
combined with the requirement of beta equilibrium, electromagnetic and
color neutrality, tends to pull apart the Fermi spheres of quarks with
different flavors~\cite{Alford:2002kj}. The mismatch between the Fermi
spheres is proportional to $M_s^2/\mu$, where 
\be\label{eq:muaverage} \mu = \frac{\mu_u+\mu_d+\mu_s}3\,, \ee is the average 
quark chemical potential. The free energy
price of having simultaneous pairing of three-flavor quark matter increases with increasing values of  $M_s^2/\mu$. 
Since the free energy gain is proportional to the CFL
gap parameter, $\Delta_{\text{CFL}}$,  if $M_s^2/\mu > c \Delta_{\text{CFL}}$, with $c$ a number of order $1$~\cite{Alford:2003fq}, a different and less symmetric CSC phase should be favored. One possibility is that the crystalline color
superconducting (CCSC) phase is realized~\cite{Alford:2000ze, Rajagopal:2006ig,Anglani:2013gfu,  Mannarelli:2014jsa}. 
In this phase quarks form Cooper pairs with nonzero total momentum, 
and there is no free energy cost proportional to $M_s^2/\mu$. 
The only free energy cost is due to the formation of counterpropagating currents; see for example the qualitative discussion in~\cite{Mannarelli:2014jsa}. 

Actually, with increasing values of $M_s^2/\mu$ various inhomogeneous CSC  phases  can be realized, because the system has many degrees of freedom~\cite{Anglani:2013gfu}. The CCSC phase   should be favored for  certain values of the chemical potential mismatch.  In reality, the CCSC phase is not one single phase but a collection of phases, characterized by their crystalline arrangements, which are favored for different values of $M_s^2/\mu$.
The Ginzburg-Landau (GL) analysis of~\cite{Rajagopal:2006ig} has shown that  
in three-flavor quark matter there are two good candidate structures that are  energetically favored for
\be
2.9 \Delta_{\text{CFL}} \lesssim \frac{M_s^2}{\mu} \lesssim 10.4 \Delta_{\text{CFL}}\,.
\ee

This range of values is certainly model dependent, moreover the GL expansion is under poor quantitative control~\cite{Anglani:2013gfu}. For this reason  we shall consider strange star models in which  both the CFL phase and the CCSC phase are realized. Since the CFL phase is expected to be favored at high densities, we shall assume that it is realized in the core of the CSO. The CCSC phase is favored at smaller densities and constitutes the crust of the CSO.   The radius, $R_{\it c}$, at which the CFL core turns into the CCSC crust will be used as a free parameter.
We shall restrict our analysis to bare strange stars~\cite{Alcock:1986hz}, meaning that we shall assume that on the top of the strange star surface there is no other layer of baryonic matter. 

Our model of  strange star resembles the typical onion structure of a standard neutron star with a solid crust and a superfluid core. It is similar to the model discussed in~\cite{Rupak:2012wk} for studying $r-$mode oscillations. In that work  the core radius was determined using a microscopic approach; instead  we treat $R_{\it c}$ as a free parameter.

One quantitative  difference between our model and standard neutron star models,  is that the CCSC phase is extremely rigid, much more rigid than the ironlike crust. The shear modulus of the energetically favored phase can be obtained studying the low energy oscillations of the condensate 
modulation~\cite{Casalbuoni:2001gt,Casalbuoni:2002pa,Casalbuoni:2002my,Mannarelli:2007bs}. In particular, the low energy expansion of the GL Lagrangian of~\cite{Mannarelli:2007bs} leads to a shear modulus
\be\label{eq:nu}
\nu \simeq \nu_0\left(\frac{\Delta}{10 \text{ MeV}}\right)^2\left(\frac{\mu}{400 \text{ MeV}}\right)^2\,,
\ee
where
\be \label{eq:nu0}
\nu_0= 2.47 \frac{\text{MeV}}{\text{fm}^3}\,, 
\ee
will be our reference value.  The reader is warned that the actual value of the shear modulus might differ from $\nu_0$ by a large amount because of the various approximations used in~\cite{Mannarelli:2007bs}. The  value of $\Delta$ is also uncertain, with reasonable values ranging
between $5$ MeV and $25$ MeV, see the discussions in~\cite{Mannarelli:2007bs,Anglani:2013gfu}.  Regarding the quark chemical potential, we shall consider the values obtained in the construction of hydrodynamically stable strange stars. The shear modulus of different crystalline structures is proportional to $\nu$, with corrections of the order unity. Since in our treatment we shall only exploit the rigidness of the CCSC phase giving the order of magnitude estimates for the various computed quantities,  the actual crystalline pattern is irrelevant for our purposes.  

Taking into account the uncertainty in the gap parameter and in the quark chemical potential,  it can be estimated that the value of $\nu$ is larger than
in conventional neutron star crust (see for example~\cite{Strohmayer}), by at least a factor of $20-1000$~\cite{Mannarelli:2007bs}. This large value of the shear modulus is due to the fact that the typical energy density  associated with  the oscillations of  the condensate 
modulation is $\mu^2 \Delta^2$, where $\Delta$ is determined by the strong interaction in the antitriplet channel. Instead, in conventional neutron star the associated energy is at the  
electromagnetic scale.  

Given the large shear modulus, one immediate consequence is that CCSC matter can sustain large deformations. 
In Refs.~\cite{Lin:2007rz, Haskell:2007sh, Lin:2013nza, Knippel:2009st, Rupak:2012wk} it has been studied the  emission of gravitational waves by various mechanisms that induce a quadrupole deformation of the CCSC structure.
See also~\cite{Andersson:2001ev} for a discussion of a different mechanism of gravitational wave emission from strange stars. 

 In the present paper we shall instead consider the electromagnetic (EM) emission by strange stars with a CCSC crust.
Since the strange star surface confines baryonic matter but allows the leakage of electrons, it follows that at the star surface there is a charge separation at the hundreds of Fermi scale~\cite{Alcock:1986hz}. Because of the large shear modulus, our model of  bare strange star can sustain large and fast torsional oscillations, leading to a periodic displacement of the surface charge. We shall see that the frequencies of torsional oscillations are of the order of MHz if the crust is hundreds of meters thick. Lower frequencies are reached if the crust is a few kilometers thick; 
GHz frequencies are reached if the crust is few centimeters thick. 

The amplitude of the oscillations at the star surface is in any case of the order of centimeters, leading to an enormous emitted radiation, of the order of $10^{41}$ erg/s, steeply increasing  for thin crusts. Thus the oscillation energy should be radiated away very efficiently, on time scales of milliseconds or even microseconds for a thin crust and of the order of hundreds of seconds for a thick crust.   More in detail,   we shall determine the frequency, the amplitude, the damping times and  the  emitted power  as a function of the various parameters that characterize the strange star. 

This paper is organized as follows. In Sec.~\ref{sec:equilibrium} we discuss spherically symmetric strange stars in hydrodynamical equilibrium. In Sec.~\ref{sec:charge} we determine the charge distribution close to the surface of the strange star.
In Sec.~\ref{sec:nonradial} we study the torsional oscillations of the strange star, estimating  the frequencies, the emitted power and the decay time  as a function of the various parameters of the model. We draw our conclusions and a possible connection with astronomical observations  in Sec.~\ref{sec:conclusions}.

\section{Equilibrium  configurations of spherically symmetric strange stars}
\label{sec:equilibrium}

For a spherically symmetric nonrotating star, 
the unperturbed background can be described by the static metric 
\be
ds^2 = g_{\mu\nu} d x^\mu d x^\nu = - e^{2 \Phi(r)} dt^2 +e^{2 \Lambda(r)} dr^2 + r^2 d\Omega^2\,.
\ee 

The relation between the function $\Lambda(r)$ and the mass
distribution $m(r)$  is given by the solution of  Einstein's
equations inside the star, namely
\be
e^{2 \Lambda(r)} = \left[1- \frac{2 m(r) G}{r} \right]^{-1},\,\,
m(r) = \int^r_0 dr' \, r'^2 \, \rho(r') \,,
\ee
where $\rho(r)$ is the energy density of the fluid.
The equilibrium structure is obtained by solving the Tolman-Oppenheimer-Volkov
(TOV) equation
\be
\frac{\partial p}{\partial r}=
-\frac{G (p+ \rho ) \left(m+4 \pi  p \, r^3\right)}{r (r-2 G \, 
   m)} \,,
\label{eq:tov}
\ee
once  the equation of state (EoS) $p(\rho)$ is specified. The star radius, $R$, is determined by the 
 boundary condition on the pressure $ p(R) =0$, simply meaning that the pressure at the surface of the star should vanish. 

The gravitational potential $\Phi$ can be found from 
\be
\frac{\partial \Phi}{\partial r} = \frac{G \left(m+4 \pi  \, p \, r^3\right)}{r (r-2 G m)}\,,
\ee
once $ p$ is derived from the solution of Eq.~\eqref{eq:tov}.    
Outside the star, for $ r > R$, defining $m(R)= M$, we have that 
\be
e^{2 \Lambda(r)} =\left(1- \frac{2 M G}{r} \right)^{-1} \, , \qquad
  e^{2 \Phi(r)} = 1- \frac{2 M G}{r} \, .
\ee

In the present work we  consider a simple strange star model, entirely composed of deconfined three-flavor quark matter in the CSC phase. The detailed form of the CSC phase is not important here, because quark pairing should account for a small variation of the  quark matter EoS. 

\begin{figure}[t!]
\includegraphics[width=9.5cm]{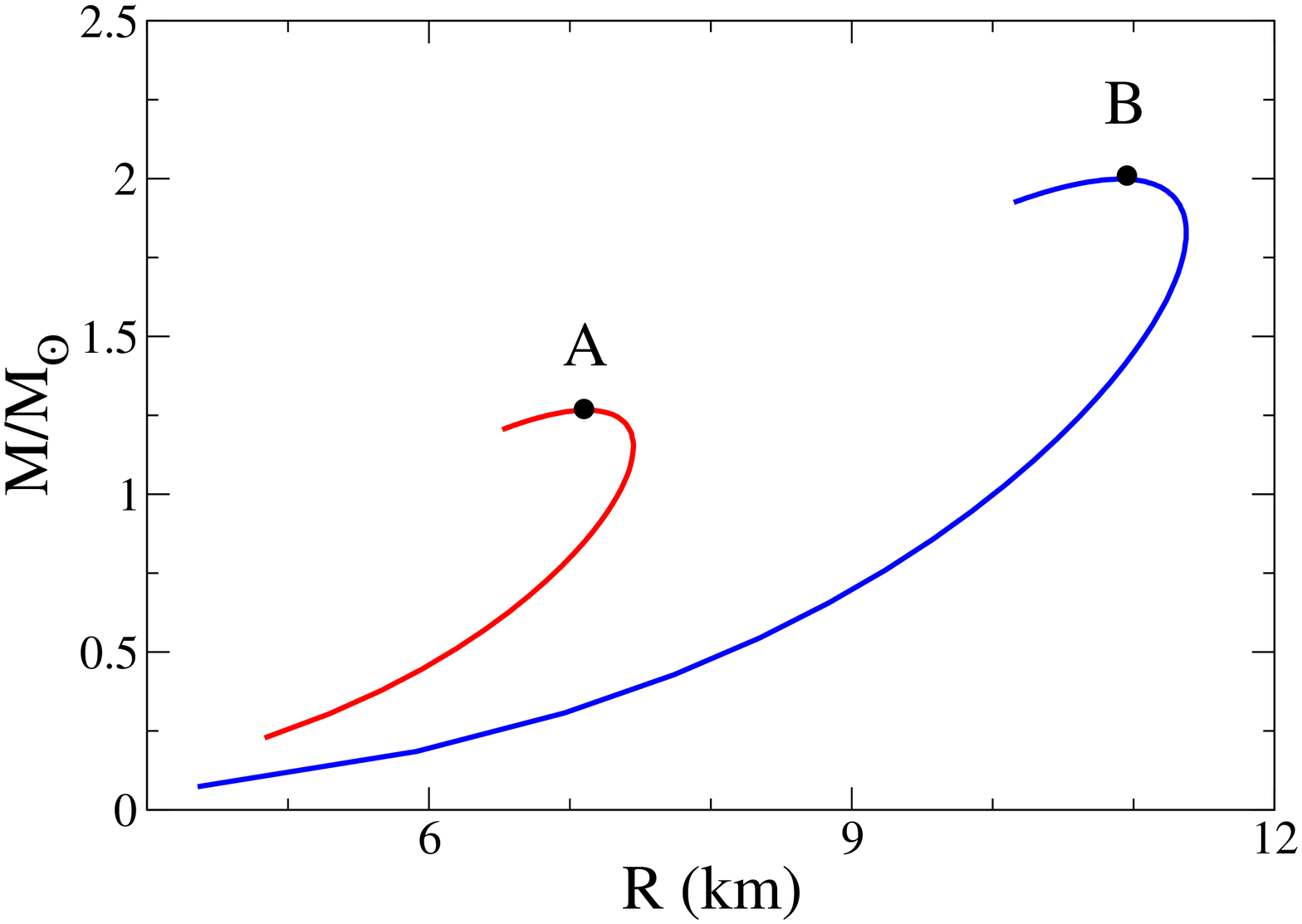}
\caption{Mass radius relation for strange stars with the EoS given in
  Eq.~\eqref{eq:EoS} for the two sets of parameters values discussed in the text. 
Changing these parameters it is possible to span a large range of
values of mass and radius. The black dots represent the equilibrium
structure assumed as reference models in this paper.
 \label{fig:MR_EOS}}
\end{figure}

Since for the range of densities attainable in compact stars QCD
perturbative calculations are not trustable, we use the general parameterization of the EoS given in \cite{Alford:2004pf}
\be\label{eq:EoS}
\Omega_{\text{QM}} = -\frac{3}{4 \pi^2} a_4 \mu^4 + \frac{3}{4 \pi^2}  a_2 \mu^2 + B_{\text{eff}}\,,
\ee
where $a_4$, $a_2$ and $B_{\text{eff}}$ are independent of the average quark
chemical potential $\mu$. This parameterization can be seen as a
Taylor expansion of the grand potential, with phenomenological
coefficients (see~\cite{Alford:2004pf}  for a discussion of the
relevant range of values of each parameter).
In order to take into account the impact of the uncertainty of these coefficients on our results,
we consider two extreme situations, namely A ($a_4=0.7$, $a_2=(200$ MeV)$^2$ and $B_{\text{eff}}=(165$ MeV)$^4$)  and 
B ($a_4=0.7$, $a_2=0$ and $B_{\text{eff}}=(145$ MeV)$^4$).
In Fig.~\ref{fig:MR_EOS} we report the mass-radius sequences obtained
solving the TOV equations using the above EoS for the two different sets
of parameters. The largest attainable mass with each set will be
our reference model, represented as a black dot in Fig.~\ref{fig:MR_EOS}. 
In detail: 
- Model A, with a total mass of $M=1.27 M_\odot$, $R\simeq
7.1$ km and  $\rho_c \simeq 5*10^{15} \text{g}/\text{cm}^3$; 
- Model B, with $M\simeq2.0 M_\odot$, $R\simeq 10.9$ km and
$\rho_c\simeq 2 *10^{15}  \text{g}/\text{cm}^3$. 
The presence of electric charge is expected to produce
corrections on masses and radii of strange quark stars at the $15\%$ and $5\%$ levels, respectively~\cite{Negreiros:2009fd}. 

\begin{figure}[h!]
\includegraphics[width=8.cm]{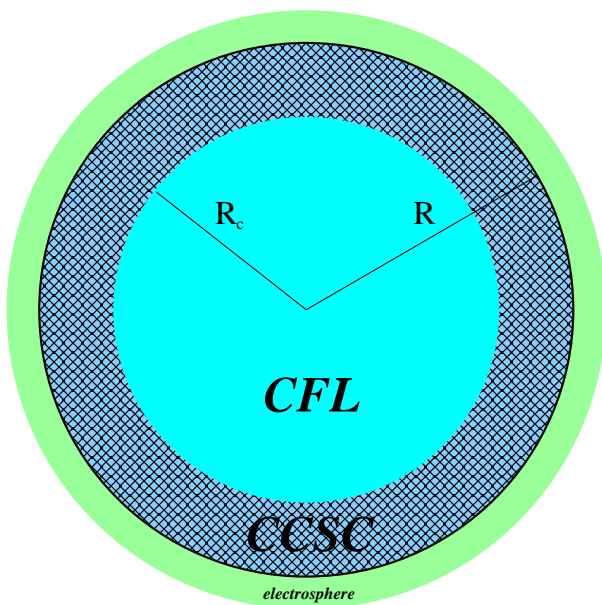}
\caption{Sketch of the  structure of the considered bare strange star model. The star {\it core} extends up to a radius $R_{\it c}$, and is made by  color-flavor locked matter.
 The  {\it crust} is made by the extremely rigid crystalline color superconducting matter.  The star radius, $R$, is determined by the solution of the TOV equation \eqref{eq:tov} with the EoS in Eq.~\eqref{eq:EoS}. We treat  the core radius, $R_{\it c}= a R$, as a free parameter. The strange star is surrounded by a cloud of electrons, the {\it electrosphere},  having a width (not in scale in the figure) of hundreds of Fermi, see Sec.~\ref{sec:charge}. 
  \label{fig:star}}
\end{figure}

In both models we assume that at a certain radial distance, $R_{\it c}=a R$ with $0 \le a \le 1$, there is a phase transition between the CFL phase and the CCSC phase.  In Fig.~\ref{fig:star} we show a pictorial description of the star structure.  Since the values of $M_s$ and of the gap parameters are unknown, it is not possible to determine from first principles the radial distance at which the CFL phase turns into the CCSC phase. For this reason  we  treat $a$ as a parameter. More in detail, in our model  we are assuming that the CFL-CCSC phase  transition does not change in an appreciable way the EoS.  Thus, our assumption is that at a given  $R_{\it c}$ the pressure of the CCSC phase and of the CFL phase are equal, but the difference between the pressure of these two phases is always small in the sense that computing the star mass and radius using only  a  CFL  EoS or only a CCSC EoS does not change the results in an appreciable way.
This is a fair approximation as far as the gap parameter in both phases are similar and  much less than the average chemical potential.   Our model is basically the model discussed in~\cite{Rupak:2012wk}, but we treat $a$ as a free parameter, whereas they compute it by a microscopic theory.

\section{Charge distribution}
\label{sec:charge}
It is very interesting to study the charge distribution close to the surface of the strange star. The reason is that quarks are confined inside the strange star by the strong interaction, while electrons can leak by a certain distance outside the star and are bound only by the electromagnetic interaction~\cite{Alcock:1986hz}. As we shall see in detail below,  electrons form an {\it electrosphere} hundreds of Fermi thick on the top of the star surface. This negative charge is compensated by a positive charge of quarks in a narrow layer, about $10$ fm thick, beneath the star surface.

Let us see how this charge separation happens. At equilibrium, the chemical potentials associated to any free particle
species must be constant, \textit{i.e.} space independent, otherwise 
particles would move to compensate the chemical potential difference. 
However, in the presence of an electrostatic potential, $\phi$, 
the density of particles can be space dependent. In the local density approximation
this fact can be taken into account defining the space dependent effective chemical potential
\be
\mu_i(\bm x) = \mu_i + e Q_i \phi(\bm x)\,,
\ee
where $Q_i$ is the charge of the species $i$, in units of the electric charge, $e$. Note that because of the weak
process  $u + d \leftrightarrow u + s$ one has $\mu_s(\bm x)=\mu_d(\bm
x)$. Note also that the average chemical potential, $\mu$, see Eq.~\eqref{eq:muaverage}, is space
independent; indeed $\mu_u(\bm x) + \mu_d(\bm x) + \mu_s(\bm x)= \mu_u + \mu_d +\mu_s=3 \mu$.

To simplify the charge distribution treatment, we shall
assume that the leading  effect of color interactions is to  provide confinement of quarks in
the interior of the star~\cite{Alcock:1986hz}. This is a good approximation if the subleading 
 effect of color interactions is the quark condensation. Indeed quark condensation is expected to produce corrections to
our results of the order $\Delta/M_s$. Since the surface of the star is in the CCSC phase  it amounts to  less than $10 \%$ corrections. We also neglect the fact that in the CCSC phase the   $U(1)_\text{em}$ is rotated to a $\tilde U(1)$, because the mixing angle between the photon and  the pertinent color field is small~\cite{Alford:1998mk, Alford:2007xm, Anglani:2013gfu}.

Therefore, we approximate  the number density of the fermionic species, $n_i$, as  a free Fermi gas, meaning that
\be
n_i(\bm x) = C_i\frac{k_{F,i}(\bm x)^3}{3 \pi^2}\,,
\label{eq:number-density}
\ee
where $C_i$ is a factor taking into account the color degrees of
freedom  and $k_{F,i}(\bm x)=\sqrt{\mu_i( \bm x)^2 - m_i^2} $ is  the
Fermi momentum, with $m_i$ the mass.  

The number density of quarks ends abruptly at the surface of the star, 
but the number density of electrons extends over  distances $r>R$,
determining the thickness of the electrosphere.
If the charge distribution varies in a region much smaller than the star radius, it is possible to approximate the 
geometry of the interface as planar. We shall assume that this is the case and then check that it is a good approximation.
For a planar interface Poisson's equation reads
\be
\frac{d^2 \phi}{d z^2} =  e \sum_i Q_i\, n_i(z) \,,
\ee
where $z$ measures the distance from the quark matter discontinuity,
located at $z=0$; the star interior corresponding to $z <0$. 

Two boundary conditions are obtained requiring  that the charge density vanishes far from the
interface. For the sake of notation we define \be V(z)=\mu_e(z)=\mu_e
- e \phi(z)\,,\ee 
and we can rewrite the  Poisson's equation as
\be\label{eq:poisson2}
\frac{d^2 V}{d z^2} = - \frac{4 \alpha_\text{em}}{3 \pi} \sum_i Q_i\, C_i\, k_{F,i}^3 \,.
\ee
Considering the weak equilibrium processes, the effective quark chemical potentials can  be written as
\be
\mu_u(z) = \mu -\frac{2}{3} V(z)\,, \,\,\, \mu_d(z) = \mu_s(z) = \mu + \frac{1}{3} V(z)\,,
\ee
and the corresponding number densities can be obtained substituting these expressions in Eq.~\eqref{eq:number-density}.
In principle the average quark chemical potential depends on the radial coordinate as well. Indeed from the EoS we can determine for any value of $r$ the function  $\mu(r)$. However, the   average quark chemical potential varies on the length scale of hundreds of meters at least, much larger than  the length scale of the quark charge distribution, which as we shall see below is of few tens of Fermi at most. Since the $z=0$ region corresponds to the star surface,  we can take  $\mu \simeq \mu_R \equiv \mu(R)$. We report in Table~\ref{table:parameter} the values of these quantities for the two considered models.

No net charge is present far from the boundary, therefore we require that 
\be n_{e}(z)_{z \to +\infty} =0\,,\ee
and that 
\be
\left[\frac{2}3 n_u(z) -\frac{1}3 n_d(z) -\frac{1}3n_s(z) - n_e (z)\right]_{z \to -\infty}=0\,. 
\ee
Neglecting the electron mass the first condition leads to $V(+\infty)= 0$; neglecting also the light quark masses the second condition leads to
\bea\label{eq:Vq}
V_q^3 &=& 2 \left(\mu_R - \frac{2}{3} V_q\right)^3 -\left(\mu_R +\frac{1}{3} V_q\right)^3\nonumber\\ &-& \left[\left(\mu_R +\frac{1}{3} V_q\right)^2- M_s^2\right]^{\frac{3}2} 
\eea
that fixes a relation among $M_s$, $\mu_R$ and $V_q=V(-\infty)$. 

Assuming that at the crust-electrosphere interface there is no surface charge, we obtain a third boundary condition  requiring that the electric field is a continuous function at $z=0$. This boundary condition can be used to obtain an expression  of the effective electron chemical potential at the surface that  depends on $V_q$ and $\mu$,  approximately given by 
\be\label{eq:V0}
V_0 = V_q -\frac{V_q^2}{2 \sqrt{3} \mu} + {\cal O}(V_q^3/\mu^2)\,.
\ee
Note that in most of the studies it is assumed that the quark distribution is constant, leading to $V_0 = 3/4 V_q$, see \cite{Alcock:1986hz}. Here we have instead used the free Fermi gas distribution for quarks, but the   quantitative result remains basically the same: the surface potential $V_0$ is smaller than $V_q$ by a not  great amount.

The Poisson's equation~\eqref{eq:poisson2} can be analytically solved for positive values of $z$ and we find  
\be\label{eq:outside_potential}
V(z) = \frac{V_0}{1+\sqrt{\frac{2 \alpha_\text{em}}{3 \pi}} V_0 z} \qquad \text{for $z>0$}\,.
\ee
Upon substituting the expression above in Eq.~\eqref{eq:number-density} (taking $C_e=1$), one readily obtains the electron distribution for positive $z$. 

In the interior of the star the Poisson's equation must be solved numerically. For negative values of $z$ we obtain by a fit of the total charge distribution, 
\be\label{eq:charge-dist}
\ \sum_i Q_i\, n_i(z)  =  b\, e^{z/d} \qquad z \le 0 \,,
\ee
where the $b$ and $d$ are two parameters describing the maximum charge density and the Debye screening length of the total charge distribution, respectively. The screening length has been computed in a different way in \cite{Alford:2011ue}, finding results analogous to ours. The values of  $V_0$ and of the fitting parameters $b$ and $d$ for the considered models and for two values of $M_s$ are reported in  Table~\ref{table:parameter}. In the interior of the star, the positive charge distribution corresponds to a shell of thickness less than  $10$ fm  peaked at $r=R$. This result is basically independent of the considered star model and of the strange quark mass. It relies on the fact that we are considering that the dominant charge carriers in the interior of the star are gapless quarks. Indeed, in the relevant CCSC phases quarks have a linear direction dependent dispersion law, see~\cite{Rajagopal:2006ig, Anglani:2013gfu}. Moreover, unpaired quarks are as present as well. For this reason the Debye screening length is approximately given by the free Fermi gas expression for three massless flavors,
\be d \simeq d_\text{Fermi gas}= \sqrt{\frac{\pi}{8 \alpha_\text{em} \mu^2} }\, \sim 5\, \text{fm} \ee
for $\mu = 300$ MeV.

The number density at the surface is approximately independent of the considered model, but depends on the chosen value of the strange quark mass. The reason is that with decreasing strange quark mass the electronic density decreases.  This effect can  also be seen from the values of the  positive surface charge density beneath the star surface
\be
Q_+ = e \sum_i  Q_i \int_0^R dr \, n_i(r) \,,
\label{eq:Q+}\ee
reported in the last column of Table~\ref{table:parameter}. 
This positive charge is balanced by the electron negative charge outside the star. The electron distribution extends outside the star for a distance approximately given by $(\sqrt{\frac{4 \alpha_\text{em}}{3 \pi} } M_s^2/\mu)^{-1}$, see Eqs.~\eqref{eq:Vq}, \eqref{eq:V0} and \eqref{eq:outside_potential},   of the order of hundreds of Fermi. Therefore, the length scales of both charge distributions are much less than the star radius. 

\begin{widetext}
\begin{center}
\begin{table}[htb!]
\begin{tabular}{|c|c|c|c|c|c|c|c|c|}
\hline Model & $\mu_{R}$ [MeV] & $\rho_{R}$ [g/cm$^3$] & $M_s $ [MeV] & $d$ [fm]& $b$ [MeV$^3$] & $V_0$ [MeV] & $Q_+$ [MeV$^3$ fm]\\
\hline A & $ 387$& $9.0 \times10^{14}$ & $150$ & $3.8$ & $4.5 \times 10^3$ & $14.2$ & $1.7 \times 10^4$ \\
\hline A & $ 387$& $9.0 \times10^{14}$ &$250$ & $3.9$ & $3.0 \times10^4$ & $37.4$ & $1.2 \times 10^5$ \\
\hline B & $ 302$& $4.1 \times10^{14}$&$150$ & $4.9$ & $5.5 \times 10^3$& $17.8$ & $2.7 \times 10^4$ \\
\hline B & $ 302$& $4.1 \times10^{14}$&$250$ & $3.3$ & $5.2 \times 10^4$ & $46.9$& $1.7 \times 10^5$\\
\hline
\end{tabular}
\caption{Values of the parameters characterizing the  surface of the two considered star models. The  second and third  columns represent the average quark chemical potential and the matter density at the surface of the star, respectively.  Considering two different values of the strange quark mass we report for each stellar model  the  parameters $d, b$ and $V_0$ characterizing the charge distributions, see Eqs.~\eqref{eq:outside_potential} and \eqref{eq:charge-dist}, and the  positive surface charge density, $Q_+$, see  Eq.~\eqref{eq:Q+}.}
\label{table:parameter}
\end{table}%
\end{center}
\end{widetext}

Note that, in principle, a  charge separation should occur as well at $R_{\it c}$,   at the CFL-CCSC interface; the reason being that the bulk CFL matter has vanishing electron chemical potential~\cite{Alford:1998mk,Alford:2007xm,Anglani:2013gfu}. However,  whatever phenomenon takes place at  the CFL-CCSC interface  should be screened by the overlying CCSC layer.

\section{Nonradial oscillations}
\label{sec:nonradial}
We now consider the possible oscillations of the above determined charge distribution.  In particular, we are interested in nonradial oscillations, which can generate an EM current at the star surface.
 
\begin{figure}[ht!]
\includegraphics[width=8.cm]{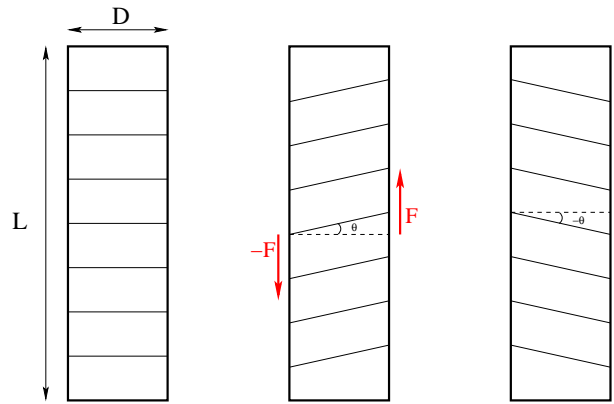}
\caption{Pictorial description of the torsional oscillations of a homogeneous two-dimensional slab with $D \ll L$. The equilibrium configuration corresponds to the one in which all the horizontal lines are parallel (left panel). A torque applied at the surfaces of the slab slightly  deforms it by an angle $\theta$ (central panel). As the applied forces vanish the slab starts to oscillate around the equilibrium configuration, reaching the configuration with deforming angle $-\theta$ (right panel). The restoring force governing the oscillation is proportional to the shear modulus, $\nu$. The applied force determines the amplitude of the oscillation; the frequency of the oscillation is proportional to $\sqrt{\nu/(\rho D^2)}$, where $\rho$ is the matter density of the slab.    
 \label{fig:slabs}}
\end{figure}

Stars have a large number of nonradial oscillations, which can be  classified as spheroidal and toroidal oscillations, see for example~\cite{1988ApJ...325..725M}.  
For definiteness,  we focus on torsional oscillations \cite{1968Natur.218.1128R, 1970Natur.225..619R, 1980ApJ...238..740H, 1983MNRAS.203..457S}, a particular class of toroidal oscillations. The  torsional oscillations are the only toroidal oscillations in nonrotating stars with a negligible magnetic field~\cite{1988ApJ...325..725M}. 
These  oscillations can be produced by acting with a  torque on a rigid structure, as shown in Fig.~\ref{fig:slabs} for a simple  rigid slab. When the applied forces are parallel to the sides of the slab they produce a deformation of the structure. As the external torque vanishes, the slab starts to oscillate around the equilibrium configuration.  The restoring force is proportional to the shear modulus and the frequency of the small amplitude oscillations is given by \be\label{eq:omegaslab} \omega \propto \frac{1}D\sqrt{\frac{\nu}{\rho}}\,,\ee where $D$ is the thickness of the slab. The slab can be thought as a local approximation of the CCSC crust, and therefore we expect that the frequency of the crust torsional oscillations has the same  qualitative dependence on $\nu$, $\rho$ and the crust thickness, $D=R-R_{\it c}$ as in Eq.~\eqref{eq:omegaslab}.

Our interest in  the torsional oscillations is clearly due to the fact that in the CCSC phase the shear modulus is extremely large and can therefore sustain large amplitude  oscillations. We shall show that for a sufficiently thin CCSC crust, the frequency of the oscillations lies in the MHz radio-frequency range, whereas for a thick CCSC crust the frequency of the oscillations lies in the hundreds of Hz range.  

Given the spherical symmetry of the nonrotating stars, it is useful to use spherical coordinates for the displacement vector \be
\xi_{nl}^r =0 \qquad \xi_{nl}^\theta = 0 \qquad \xi_{nl}^{\phi} = \frac{W_{nl}(r)}{r \sin\theta} \frac{\partial P_l(\cos \theta)}{\partial \theta} e^{i \omega_{nl} t}\,,
\ee
where $l$ is the angular momentum and $n$ is the principal quantum number indicating the number of nodes.
In the following we shall study these oscillations in the Newtonian approximation. We expect that a treatment with full general relativity (GR) should give correction factors of order unity. Given the large uncertainties of the various parameters of the model, it seems appropriate to neglect GR corrections. We shall investigate the GR corrections in  a future work.

In the Newtonian limit the  velocity perturbation\footnote{Eulerian and Lagrangian perturbations are identical for toroidal oscillations~\cite{1988ApJ...325..725M}.} is given by
\be
\delta \bm u_{nl} =  i \omega_{nl} \bm \xi_{nl} \,,
\ee
and the amplitude of the horizontal oscillation satisfies the following differential equation
\bea
\omega_{nl}^2 W_{nl} &=& \frac{\nu}{\rho} \left[-\frac{1}{\nu} \frac{d \nu}{d r} \left(\frac{d W_{nl}}{ d r}-\frac{W_{nl}}{  r}\right) -\frac{1}{r^2}\frac{d}{dr}\left( r^2\frac{d W_{nl}}{ d r} \right) \right. \nonumber\\ &+& \left. \frac{l(l+1)}{r^2}W_{nl} \right]\,.
\eea
The torsional oscillation extends in the CCSC crust and disappears suddenly in both  the CFL phase and  the electrosphere.  The  displacement is discontinuous  at these interfaces because 
both the CFL and the electrosphere have a vanishing shear modulus.

We shall simplify the discussion assuming that 
the quark matter in the CCSC crust has constant density,  $\rho\simeq \rho_R=\rho(R)$. 
This is a good approximation because in all the considered   cases correspond to a CCSC crust a few kilometers thick at most. In any case, the density of quark matter in strange stars does not sharply change, because strange stars are self-bound CSOs.  We shall as well neglect the radial dependence of the shear modulus and take it as a constant.
In this way  the displacement satisfies the following  differential equation:
\be\label{eq:differential1}
\frac{d^2 W_{nl}}{ d r^2} + \frac{2}{r} \frac{d W_{nl}}{ d r}+ \left(\frac{\omega_{nl}^2}{v_s^2} - \frac{l(l+1)}{r^2}\right)W_{nl}=0\,,
\ee
where $v_s= \sqrt{\nu/\rho_R}$ is the shear wave velocity.
It is useful to switch to the adimensional variable $y= \omega_{nl} r/v_s$, and to define
$W_{nl}= U_{nl}/y$. In this way the  above differential equation can be written as
\be\label{eq:differential2}
U_{nl}''(y)  + \left(1 - \frac{l(l+1)}{y^2}\right) U_{nl}(y) =0 \,.
\ee
The solution of this equation can be expressed as a sum of spherical Bessel and Neumann functions
\be\label{eq:solution}
U_{nl}(y) =  c_1 j_{l}(y) + c_2 n_l(y)\,.
\ee
After an initial stage in which the crust has been excited by some external agency, we assume that there is no torque acting on both interfaces of the crust. This corresponds to assuming the  no-traction condition \cite{1988ApJ...325..725M}, leading to 
\be\label{eq:BC}
U_{nl}'(y_1) = 2\frac{U_{nl}(y_1)}{y_1} \qquad U_{nl}'(y_2) = 2\frac{U_{nl}(y_2)}{y_2}  \,,
\ee
where $y_2=\omega_{nl} R/v_s$ corresponds to the CCSC-electrosphere interface and $y_1 =\omega_{nl} aR/v_s = a y_2$ corresponds to the CFL-CCSC interface. One of the two  conditions can be used to eliminate one of the two coefficients in Eq.~\eqref{eq:solution}. The other  condition determines the quantized frequencies. For definiteness, we shall hereafter assume that the only excited mode is the one  with $l=1$ and with one node\footnote{The mode with $n=1$ is the first nontrivial mode for $l=1$. Indeed the mode with no nodes $(n=0) $ corresponds to a global rotation of the star. }, $n=1$. In this case we find that for $a \gtrsim 0.3$ one can use the approximate functions
\be
y_2 \simeq \frac{\pi}{1-a} \qquad y_1 \simeq \frac{a \pi}{1-a}\,,
\ee
and the corresponding oscillation frequency  is given by
\be\label{eq:w1}
\omega_{11} \simeq 0.06  \left( \frac{\nu}{\nu_0} \right)^{1/2}\!\!\!\left( \frac{\delta R}{1 \text{km}} \right)^{-1} \left(\frac{\rho_R}{\rho_0} \right)^{-1/2} \text{MHz}\,,
\ee 
where $\delta R= (1-a)R$ and $\rho_0=10^{15} \text{g/cm}^{3}$. Note that for each set of parameters the equation above gives the smaller attainable frequency. Radial
overtones,  having higher frequencies, could  as well be excited by the external agency triggering the oscillation of the crust.

As in the simple example of the slab (see the caption of Fig.~\ref{fig:slabs}) the amplitude of the oscillations is determined by the external agency, which  fixes  the amount of energy of each mode.  The frequency of the oscillations is proportional to the shear velocity divided by the crust width.

\begin{figure}[t!]
\includegraphics[width=8.cm]{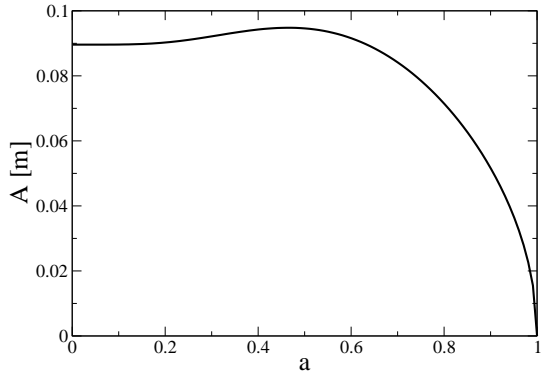}
\caption{Function determining the horizontal displacement at the surface of the star associated with   the considered torsional oscillation, see Eq.~\eqref{eq:W11}.
 \label{fig:amplitude}}
\end{figure}

For definiteness, we shall assume that  a fraction $\alpha$ of the energy of a glitch excites the $l=1, n=1$ mode; thus
\be\label{eq:glitch}
\alpha E_{\text{glitch}} = \frac{\rho_R}{2}  \int  |\delta \bm u_{11}|^2 dV \,,
\ee
where  we shall consider as a reference value $E_{\text{glitch}}^{\text{Vela}} = 3\times10^{-12} M_\odot$ as estimated for the giant Vela glitches. Less energetic glitches, as for the Crab, simply correspond to  smaller values of $\alpha$.   Of particular relevance for us is the amplitude of the oscillation at the star surface, because it  determines the displacement of the quark electric charge. The displacement does in general depend on the thickness of the crust, on the shear modulus and can be expressed as
\be\label{eq:W11}
W_{11}(R) = A(a)  \left( \frac{\nu}{\nu_0}\right)^{-1/2} \left(\frac{R}{10 \text{km}}\right)^{-1/2}\left(\frac{\alpha E_{\text{glitch}}}{E_{\text{glitch}}^{\text{Vela}} }\right)^{1/2}\,,\,
\ee
where $A(a)$ is reported in Fig.\ref{fig:amplitude}. 

The amplitude of the  oscillations is in general quite large, of the order of centimeters, except for $a \simeq 1$. Indeed,  the amplitude vanishes for $a=1$, because this case corresponds to a star completely made of CFL matter. In the following we shall always consider the case in which the crust has a macroscopic extension, much larger than the extension of the positive charge distribution. Notice that considering  a perturbation that excites  more modes  does not qualitatively change the above results. The only effect is a distribution of the total energy among modes with higher frequency. 

The  fluctuation of the current density induced by periodic horizontal displacement associated with the torsional oscillation is given by 
$\delta{\bm J}=  e \sum_i Q_i\, n_i(z) {\delta\bm u_{11}}$.  The  EM vector field  in a point outside the source is given by
\begin{equation}
\delta{\bf A}({\bf r}, t)=\int \frac{\delta{\bf J}({\bf{r'}},
  t_R)}{|{\bf r}-{\bf r'}|}d V'\,,
\end{equation}
where $t_R=t-|\bm r - \bm r'|$. We estimate the emitted power by considering the moving electric charges as an oscillating magnetic  dipole. In the far field approximation ($|\bm r| \gg |\bm r'|$), and for a coherent emission ($\omega_{11} \ll 1/ |\bm r'|$), we obtain that
\bea
P &=&  \frac{e^2 \pi^4 \omega_{11}^6}{6} \left(  \sum_i Q_i \int_0^R dr \, n_i(r)  r^3  W_{11}(r)\right)^2 \sin^2(\omega_{11} t) \nonumber\\ &\simeq&  \frac{\pi^4 (\omega_{11} R)^6}{6} W_{11}(R)^2 Q_+^2 \sin^2(\omega_{11} t)\,,
\label{eq:P}\eea
where $Q_+$ is given in Eq.~\eqref{eq:Q+}.
An approximate value of the radiated power is given by
\begin{widetext}
\be\label{eq:Pa}
P(a) \simeq 6.4 \times 10^{41} \left(\frac{y_2(a)^6 A^2(a)}{y_2(0)^6 A^2(0)}\right)  \left(\frac{\nu}{\nu_0} \right)^2  \left(\frac{\rho_R}{\rho_0} \right)^{-3} \left(\frac{R}{10 \text{km}}\right)^{-1} \left(\frac{\alpha E_{\text{glitch}}}{E_{\text{glitch}}^{\text{Vela}} }\right) \left( \frac{Q_+}{Q}\right)^2 \text{erg/s}\,,
\ee
\end{widetext}
where we have averaged over time and considered as a reference value for the surface charge density $Q= 10^5$ MeV$^3$fm. The values of $Q_+$ for the two considered models and for two different values of $M_s$ are reported in Table \ref{table:parameter}.  The radiated power increases with increasing $a$, meaning that the thinner the crust, the larger the radiated power (as far as the crust remains larger than the region in which there is a positive electric charge). For example considering $a=0.9$ we obtain a radiated power of about  $10^{45}$ erg/s.
The radiated power increases because the oscillation frequency increases with increasing $a$. It certainly happens that for $a \to 1$ the amplitude of the oscillation decreases, see Fig.~\ref{fig:amplitude}, but it does not decrease fast enough to compensate for the increase of the frequency, and thus the  product $y_2(a)^6 A^2(a)$ in Eq.~\eqref{eq:Pa} increases with increasing $a$.

\begin{figure}[b!]
\includegraphics[width=8.cm]{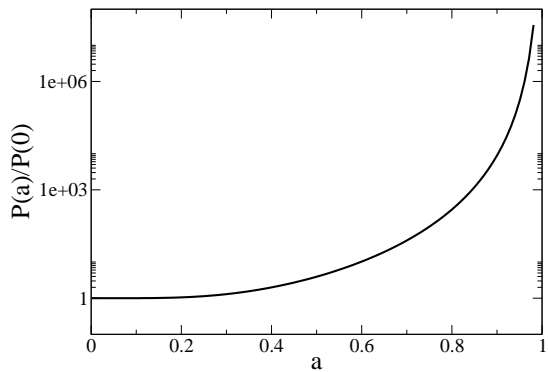}
\caption{Radiated power, see Eq.~\eqref{eq:Pa}, as a function of  the ratio between the core radius and the star radius, $a$.  The radiated power is normalized to the value for $a=0$, corresponding to  a strange star with a rigid crust extending from the surface down to the center of the star.
 \label{fig:radiated power}}
\end{figure}

\begin{table}[ht!]
\begin{center}
\begin{tabular}{|c|c|c|c|c|}
\hline Model & $ M_s $ [MeV] & $\Delta$ [MeV] & $\tau_{a=0.9}$[s]&$\tau_{a=0.1}$[s] \\
\hline A &   $150$ &  $5$ & $2.0 \times 10^{-1}$ & $1.7 \times 10^{3}$ \\
\hline A &   $150$ &  $25$ & $3.1 \times 10^{-4}$ & $2.8\times 10^{1}$\\
\hline A &   $250$ &  $5$ & $4.2 \times 10^{-3}$ & $3.7 \times 10^{1}$\\
\hline A &   $250$ &  $25$ & $6.7 \times 10^{-6}$ & $5.9 \times 10^{-2}$\\
\hline B &   $150$ &  $5$ & $3.3 \times 10^{-1} $ & $2.9 \times 10^{3}$\\
\hline B &   $150$ &  $25$ & $5.2 \times 10^{-4}  $& $4.6 \times 10^{0}$\\
\hline B &   $250$ &  $5$ & $8.2 \times 10^{-3}$& $7.3 \times 10^{1}$ \\
\hline B &   $250$ &  $25$ & $1.3 \times 10^{-5}$& $1.2 \times 10^{-2}$\\
\hline
\end{tabular}
\end{center}
\caption{Approximate values of the damping times for the two considered models for two different values of the strange quark mass and of the gap parameter. In the third column we have assumed that $a=0.9$, meaning that the CCSC crust is  about $0.7$ km ($1.1$ km)  thick for Model A (Model B). In the fourth column we have taken $a=0.1$, meaning that the CCSC crust is about $6.4$ km ($9.8$ km) thick for Model A (Model B).}
\label{table:damping}
\end{table}
Approximate values of the damping time, $\tau$, for the two considered models and for different values of $M_s$ and $\Delta$ are reported in Table \ref{table:damping}, for two values of $a$. The damping time is computed simply dividing the energy of the oscillation for the corresponding emitted power. This certainly gives a rough, order of magnitude, estimate of the time needed for emitting all the energy. We find that $\tau$ decreases  with increasing values of  $M_s$ and/or $\Delta$. The reason is that with increasing values of $M_s$ the positive charge close to the star surface increases, see Eq.~\eqref{eq:charge-dist} and Table \ref{table:parameter}.  With increasing values of $\Delta$ the shear modulus increases, see Eq.~\eqref{eq:nu}, and therefore the frequency of the oscillations increases. Since  $P \propto Q_+^2 \omega_{11}^6$,  the dependence on $\Delta$ is particularly strong, see Eq.~\eqref{eq:w1} and   Eq.~\eqref{eq:nu}. \\

So far we have assumed that the electrosphere does not screen the radiated photons.  
However, it is conceivable that a sizable fraction of the emitted energy will be scattered by the electrosphere. The detailed modelization of the electrosphere and of the  interaction with the photons emitted by the positive charged quark layer is a nontrivial problem, see for example~\cite{Cheng:2003hv, PicancoNegreiros:2010uc, Zakharov:2010yz, Rupak:2012wk}. We shall 
 estimate the absorbed fraction considering a completely degenerate electron gas at small temperature~\cite{Cheng:2003hv}. The absorption is due to the Thomson scattering of photons off degenerate electrons and leads to an exponential reduction of the emitted power. Assuming that electrons are degenerate and considering that temperature corrections, of order $T/m_e$, can be neglected, one finds that the intensity of the emitted radiation is suppressed  by~\cite{Cheng:2003hv} 
\be\label{eq:etaeff}
\eta = \exp\left(-\frac{1.129 }{6 \pi^{2}} \sqrt{\frac{3 \pi}{2\alpha_\text{em}}} \sigma_{\it T} V_0^2\right)\,,
\ee
where $\sigma_{\it T}=8\pi/3 (\alpha_\text{em}^2/m_e^2)$ is the total Thomson cross section. 
For the considered values of the surface potential, see Table~\ref{table:parameter}, we obtain  suppression factors of order $0.1$.

\section{Conclusions}
\label{sec:conclusions}

We have discussed two very simple strange star models entirely composed  of deconfined color superconducting  matter. Our star models are very similar to those proposed in~\cite{Rupak:2012wk} for the discussion of $r-$mode oscillations.  We assume that the star core is composed by CFL matter and there is a crust of rigid CCSC quark matter. The size of the star crust is unknown, because it depends on the detailed values of the strange quark mass and of the gap parameters, which are very uncertain. For this reason we have treated the ratio between the core radius and the star radius as a free parameter.

In our treatment of the crust we have  determined the equilibrium charge configuration, in Sec.~\ref{sec:charge},  using the free Fermi gas  distributions but at the same time we have considered  CCSC matter, in Sec.~\ref{sec:nonradial}, as a rigid crystalline structure. These two facts may seem to be in contradiction. However, in the relevant  crystalline phases, quarks close to the Fermi sphere  have  a linear dispersion law~\cite{Casalbuoni:2003sa, Anglani:2013gfu}, which indeed mimics the behavior of free quarks. The  effect of the condensate is to induce a direction dependent Fermi velocity. Moreover, not all quarks on the top of the Fermi sphere are paired. For these reasons we have assumed that the EM properties of the CCSC phase are similar to those of unpaired quark matter.   What is rigid is the modulation of the underlying quark condensate that can be seen as the structure on the top of which quarks propagate. This treatment of the EM properties of the CCSC phase  is in our opinion an educated assumption.  A detailed study of the EM properties of the CCSC phase is necessary to substantiate this approach.

The  discussion of the quark matter surface can certainly be improved including condensation effects, following for example the discussion in~\cite{Alford:2001zr}, or viscous damping below the star crust as in~\cite{Lindblom:2000gu}. Moreover it would be interesting to study whether strangelet crystals could form on the star surface~\cite{Jaikumar:2005ne}. One should investigate whether these strangelet nuggets might coexist with the CCSC phase, presumably assuming that  the surface tension of quark matter is not too large, eventually leading to  a drastic reduction of the surface charge density.

In our simple treatment, we estimate the emitted energy of the torsional oscillations  using an oscillating magnetic dipole. The emitted power is extremely large, for stars with a small CFL core it is of the order of $10^{41} \eta$ erg/s, where $\eta$ is a screening factor due to the presence of the electrosphere, estimated in Eq.~\eqref{eq:etaeff}. The emitted power steeply increases with increasing values of $a$, see Fig.~\ref{fig:radiated power}, meaning that stars with a large CFL core and a thin CCSC crust would probably emit all the oscillation energy in  milliseconds. For a sufficiently thin  crust, say hundreds of meters thick, we expect that the emission  is at the MHz frequency.  

Given the large  emitted power, it is  tempting to compare our results with  the  most powerful observed  EM emissions. The  radio bursts observed from Rotating Radio Transients have a duration of few milliseconds and the associated flux energy is extremely large~\cite{McLaughlin:2005eq}. 
However, the observed frequencies are of the order of GHz, see~\cite{Lorimer:2007qn, Thornton:2013iua, 2013Sci...341...40C}, whereas we found that the  frequencies associated with the $n=1, l=1$ mode of a star with a $1$ km CCSC crust   are of the order of tens of kHz. If the crust is thinner, say a few centimeters thick, then GHz frequencies can be attained, but in this case the emitted power, estimated by Eq.~\eqref{eq:Pa}, is extremely large and the damping time should be much less than the observed milliseconds. A loophole  might be that by Eq.~\eqref{eq:Pa}  we are overestimating the emitted power by orders of magnitude. The reason is that in Eq.~\eqref{eq:Pa} we are 
using the coherent emission approximation, which conceivably breaks down at such large frequencies.  Moreover,
we are assuming the presence of a net positive charge, with electrons only providing a screen for the emitted power. A more refined treatment should include the effect of the star magnetic field which might strongly couple  the oscillation of the star and of the electrosphere. Therefore, future work in this direction is needed to clarify how the presented discussion of the emitted EM power  changes in the presence of strong magnetic fields, see for example \cite{Glampedakis:2006apa, Levin:2006qd}. 
 A different  possibility is that  there exists a mechanism for exciting predominantly modes with higher angular momentum and/or higher principal quantum numbers. In this case, frequencies larger than  tens of kHz could be reached  for stars with a thick crust, as well.

Different  powerful phenomena of great interest are giant magnetar x-rays flares~\cite{Strohmayer:2005ks}. The observation of these flares has posed a challenge to strange stars with no crust~\cite{Watts:2006hk}. The standard explanation of these flares is indeed related with the seismic vibrations of the crust triggered by a starquake. Typical frequencies are of the order of hundreds of Hz  at most and the emitted luminosities is of the order of $10^{44} - 10^{46}$ erg/s. The measured decaying time is of order of minutes. In our model,  oscillations of hundreds of  Hz can be reached only if the shear modulus is sufficiently small, of the order of $\nu_0 10^{-4}$, making it comparable with standard  nuclear crusts,   and if the CCSC crust is  sufficiently thick, say of the order of a few kilometers, meaning that $a \sim 0.1$. For these small values of $a$ the damping time can be  of the order of hundreds of seconds, see the last column in Table~\ref{table:damping}. Basically, our bare strange star model has frequency and decay times compatible with magnetar flares  only if it has  the same structure of  a standard neutron star. The caveat is that these flares are observed in magnetars, which are CSOs expected to have a large magnetic field.  In our simple treatment we have neglected the effect of the background magnetic field, which however in the case of magnetars could be sizable. 

We have neglected the effect of the temperature, as well. Although temperature effects are negligible for  strange stars older than $\sim10$ s \cite{Page:2002bj}, when the temperature has dropped below the MeV scale, it would be interesting to see what is the effect of a large, say $\sim 10$ MeV temperature, see for example the discussion in \cite{Cheng:2003hv, Usov:2004kj}. Unfortunately, the shear modulus has only been computed at vanishing temperature~\cite{Mannarelli:2007bs}. A detailed study of the temperature dependence  of the shear modulus and of the response of the CCSC structure to the temperature  is needed to ascertain the correct temperature dependence of  the torsional oscillations.  However, let us assume that the CCSC structure has already formed when the temperature is of the order of few MeV,  and that it responds to the temperature as a standard material, meaning that with increasing temperature the shear modulus  decreases.  From Eq.~\eqref{eq:w1} it follows that the frequency of the torsional oscillation decreases, and, from Eq.~\eqref{eq:W11}, that the amplitude increases. Moreover, an increasing temperature should lead  to an  increase of the number densities of the light quarks and electrons, leading to  a larger $Q_+$.   The overall effect on the radiated power is not obvious, because in   Eq.~\eqref{eq:P} we have that  $\omega_{11}$ decreases, but $W_{11}$ and $Q_+$ increase; therefore a careful study of the various contributions is necessary. Regarding  the emission mechanisms of the electrosphere,  one should consider the various processes that become relevant at nonvanishing temperature. In particular,  $e^+e^-$ production  is believed to be the dominant process at high  temperature~\cite{Usov:1997ff, Usov:2001sw, Aksenov:2003kh, Aksenov:2003vy}. 

Finally, note that in the evaluation of the horizontal displacement  one should include  the radial dependence of the shear modulus and of the matter  densities as well as  GR corrections, which are expected to be small, but should nevertheless be considered for a more refined  study.

\acknowledgments
We thank M.~Alford and I.~Bombaci  for insightful discussions.

\bibliographystyle{h-physrev4}

\end{document}